\documentclass[%
 aip,
 amsmath,amssymb,
 reprint,%
]{revtex4-2}
\usepackage{graphicx}
\usepackage{dcolumn}
\usepackage{bm}

\usepackage[utf8]{inputenc}
\usepackage[T1]{fontenc}
\usepackage{mathptmx}

\begin{document}


\title{Development of a Pulsed VUV Light Source \\With Adjustable Intensity}

\author{A.D.~McDonald}
\email{austin.mcdonald@uta.edu}
\affiliation{ Department of Physics, University of Texas at Arlington, Arlington, TX 76019, USA }%
\affiliation{ Department of Physics, Harvard University, Cambridge, MA 02138, USA }%

\author{M. Febbraro}%
\email{febbraromt@ornl.gov}
\affiliation{Physics Division, Oak Ridge National Laboratory, Oak Ridge, TN 37831, USA}%

\author{J.~Asaadi}
\affiliation{ Department of Physics, University of Texas at Arlington, Arlington, TX 76019, USA }%

\author{C.C~Havener}
\affiliation{Physics Division, Oak Ridge National Laboratory, Oak Ridge, TN 37831, USA}%

\date{\today}

\begin{abstract}
This paper describes the development of a pulsed light source using the discharge from an electrode in a medium of various noble gases. This source can be used to aid in the characterization and testing of new vacuum-ultraviolet (VUV) sensitive light detection devices. The source includes a novel spark driver circuit, a spark chamber into which different noble gases can be introduced, and an optical attenuation cell capable of being filled with different gases to allow for the attenuation of the pulsed light down to single photon levels. We describe the construction, calibration, and characterization of this device deployed at a dedicated light detection test stand at Oak Ridge National Laboratory.
\end{abstract}

\maketitle

\section{\label{sec:level1}Introduction}

Vacuum-ultraviolet (VUV) photon detectors capable of directly detecting VUV photons at low photon (and ultimately single photon) flux have been an area of active research for the past few decades. These VUV photon detectors come in many different kinds including ones based on scintillator, novel photoconductors,  photomultiplier tubes (PMTs), semiconductors, and gas based detectors. The article from Zheng et al. \cite{Zheng2020} summarizes the recent progress across the broad class of detectors and their various applications.

Of particular interest in the field of High Energy Physics (HEP) are VUV photon detectors capable of detecting light from gas and liquid noble gas detectors. Noble element detectors have become ubiquitous in searches for dark matter and as a medium for detection of neutrino-nucleus interactions \cite{app11062455, BAUDIS201450}. The detection of the scintillation light arising from  particle interactions is a key component of these detectors and thus necessitate the need for devices sensitive to this scintillation light. The scintillation light for the most common noble element detectors (Argon \& Xenon) sits in the VUV spectrum of 128~nm – 175~nm (9.7 eV - 7.1 eV). 

\begin{figure}[t]
\centering
\includegraphics[width=1.0\linewidth]{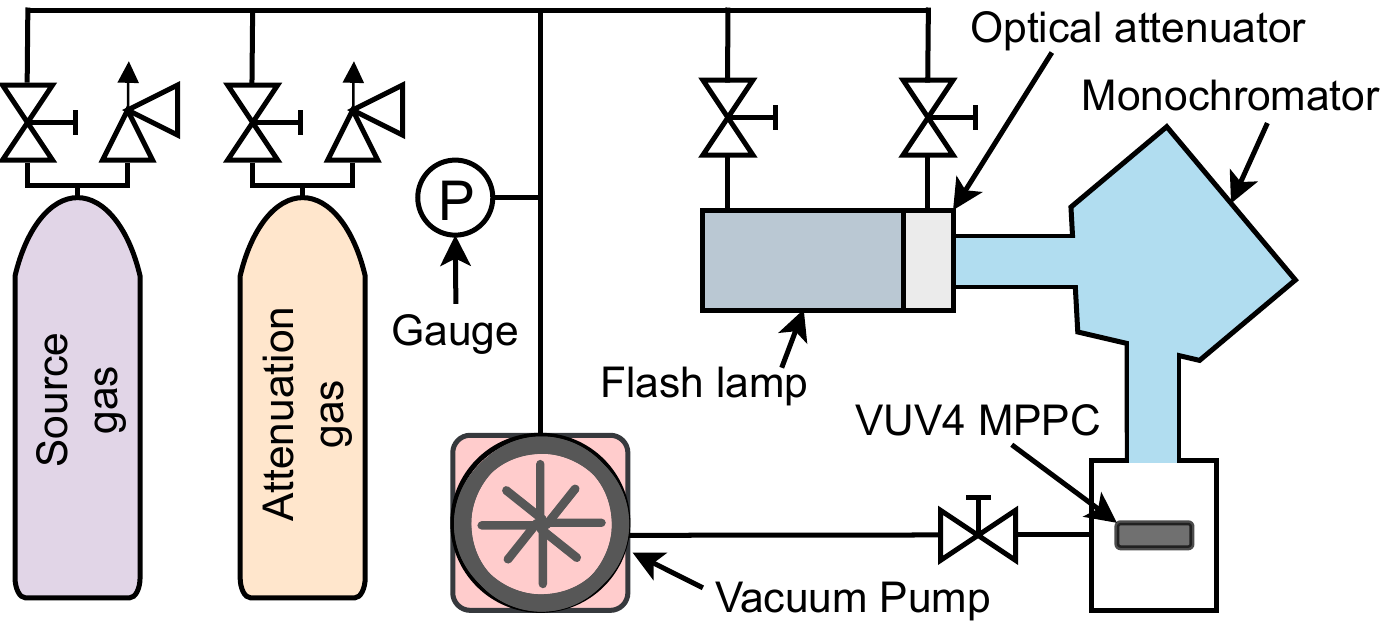}
\caption{Schematic view of the overall experimental setup. The gas handling for the flash lamp and optical attenuator can be seen and they share a common line which is evacuated when not in use. The pressure in the gas lines/volumes is measured via the gauge when filling. The monochromator and VUV4-MPPC are kept at vacuum continuously. Light from the lamp passes though the attenuator cavity and enters the monochromator where a chosen wavelength is then sent to the VUV4-MPPC. }
\label{fig:Overall}
\end{figure}

A challenge for devices developed to be sensitive to VUV photons in the range produced by noble element detectors is finding a calibration source capable of: i) producing and transmitting photons in the VUV (128~nm - 200~nm), ii) has a controllable intensity capable of reliably obtaining single photon transmission, and iii) can produce a pulsed source of these photons such that the readout of the VUV photon detector can be correlated with the light source and the characteristics of the photodetector can be measured using the pulsed source.

Conventionally, when the need for a VUV light source is called upon, deuterium lamps with a magnesium fluoride window can provide a constant source of VUV light down to 115~nm \cite{deutlamp}. Such a source allows you characterize the overall responsiveness and quantum efficiency of the device under test, but not the response the device will have to a pulse of VUV light (as would be seen in an actual detector). Another commonly employed calibration source is the Xenon flash lamp, such as those found in references \cite{xenonflashlam}. These offer a pulsed light typically microseconds in length with kiloHertz repetition rate. However, the spectrum produced is broad and typically has a cutoff at 200~nm based on the optical coupling elements design. Moreover, the intensity of the light is often very high requiring non-trival attenuation at wavelengths which are difficult to attenuate.

\begin{figure*}[t]
\centering
\includegraphics[width=\textwidth]{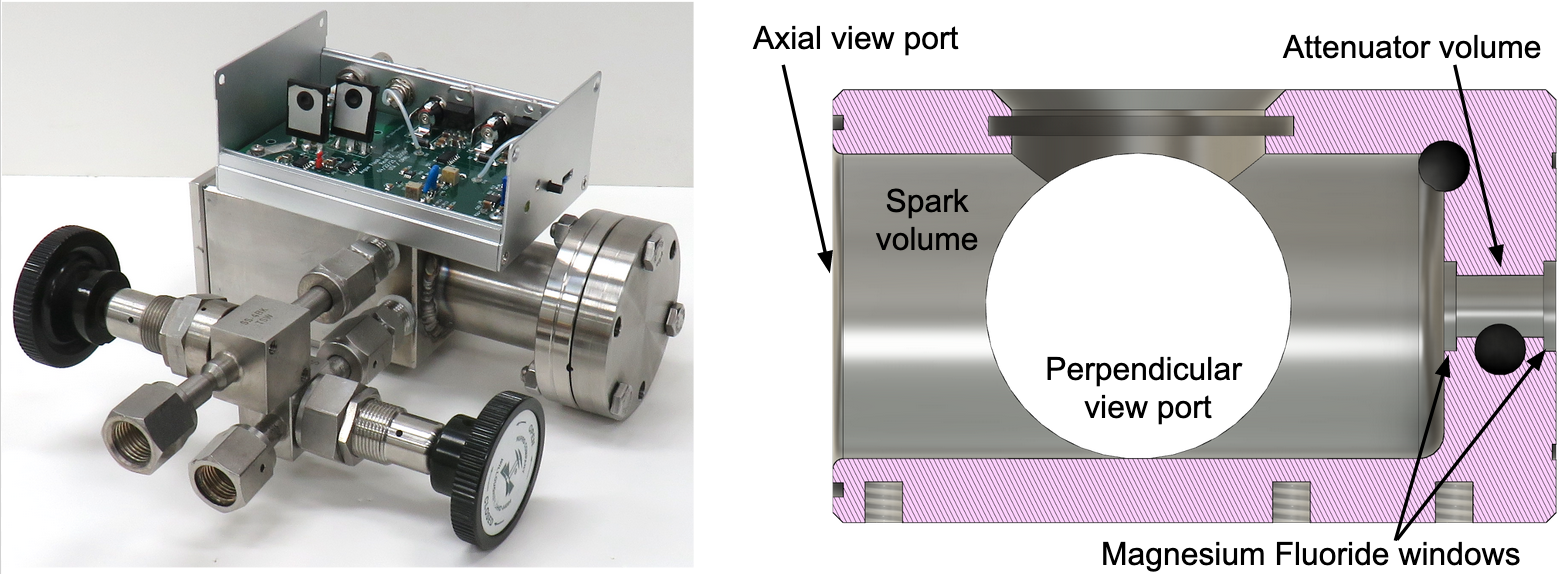}
\caption{Left: View of the completed pulsed source assembly with the electronics mounted to the top and the cover removed. The VCR valves are kept affixed to the body to reduce the gas volume and number of connections. Right: A cross sectional view of the engineering model where the internal details can be viewed. Notably, where the gas connections are made to the inner volumes, how the epoxy seal is formed around the discharge electrodes, and how the optical attenuator is formed via the two MgF$_{2}$ windows epoxied into the counter bores.}
\label{fig:Lamp}
\end{figure*}

To better address the characterization and testing of new VUV sensitive light detection devices, a pulsed light source using the discharge from an electrode in a medium of noble element gases has been designed and commissioned. This device includes an optical attenuation cell capable of being filled with different gases to allow for the attenuation of the pulsed VUV light down to single photon levels. The capability of producing both pulsed VUV photons and the ability to attenuate these to low photon count for characterization of a device under test offers a new and novel tool for the prototyping of VUV sensitive light detection devices. In this paper we describe the details this pulsed light source. Section \ref{sec:apparatus} describes the experimental test setup, the mechanical aspects of the assembly, and the details of the electronics which allow for a fast discharge and configurable pulse width. Section \ref{sec:Methodology} describes the operation, calibration, and analysis of the data taken with the pulsed light source. Finally, Section \ref{sec:results} presents the results and offers a discussion on the performance of the device.

\section{\label{sec:apparatus}Apparatus}

\begin{figure*}[t]
    \centering
    \includegraphics[width=1.0\textwidth]{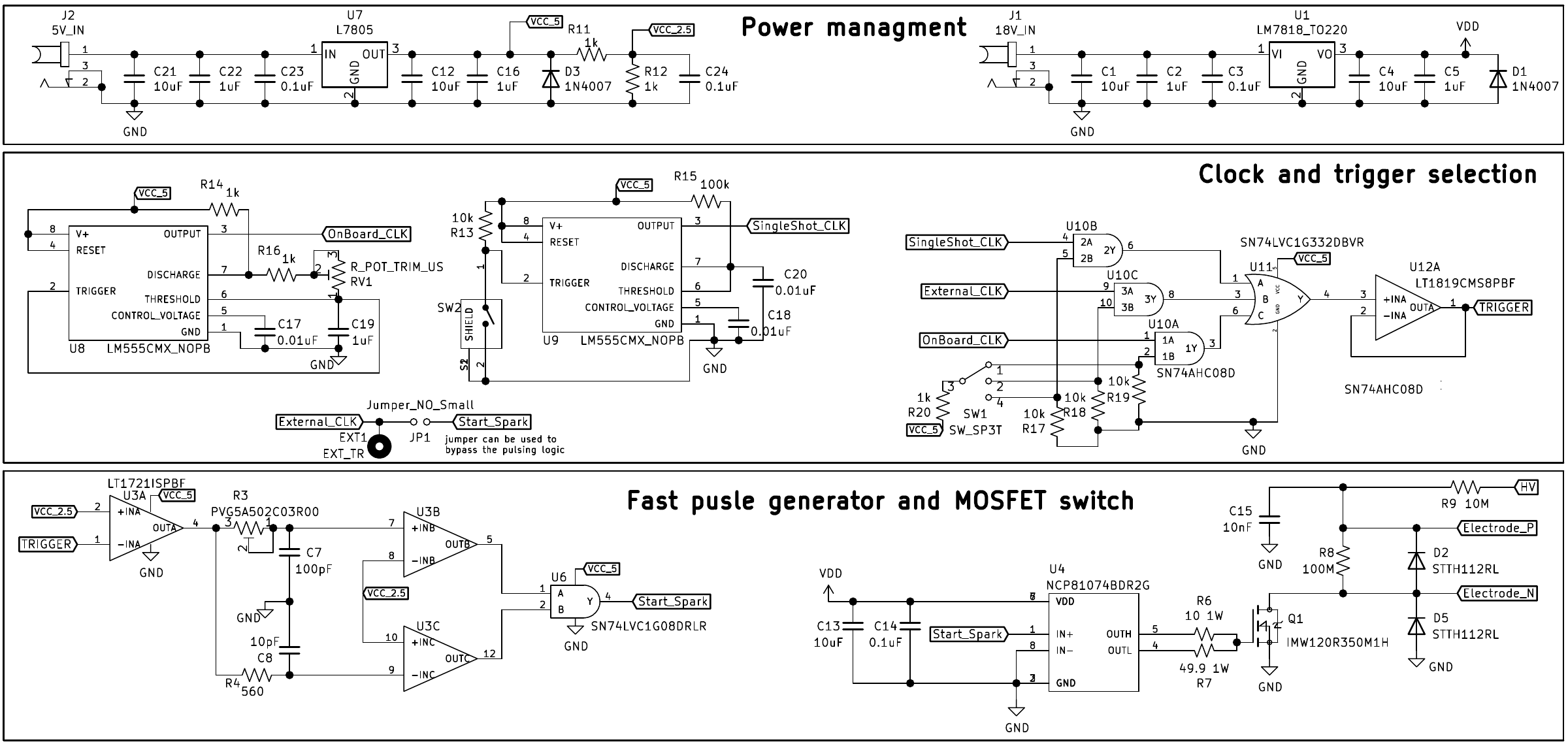}
    \caption{Overall view of the discharge electronics. The top block shows the power management which supplies a steady 5~V and 2.5~V from a 7805 regulator for powering the logic components. A 7818 regulator supplies a constant 18V for the MOSFET switch (NCP81074BDR2G), this was done to operate the MOSFET (IMW120R350M1H) at peak performance. The middle block contains the clock and triggering components. There are two 555 timers, one is in a monostable configuration, the other is an astable configuration. The monostable 555 timer is triggered via a push button and can trigger the lamp on demand. The astable 555 has an adjustable resistance which in turn adjust the the 555's frequency, yielding a tuneable on board clock. An external clock can also be inputted via a BNC connection. The clock signals are then AND'ed with a selector switch and the three outputs are OR'ed together which produces a single trigger signal. The bottom block contains the fast pulsing circuit witch is described in an application note \cite{williams2004signal} along with the MOSFET switch and high voltage components. The switch is configured according to the data sheet, the small resistances on the output were selected for a fast turn on performance. The high voltage components were selected to have a moderate charging time (RC=1~ms), a large resistor is mounted in between the electrodes to maintain a path to ground, and the diodes allow for transient protection.}
    \label{fig:schematic}
\end{figure*}

The experimental setup developed to test the performance of the pulsed light source is shown schematically in Figure \ref{fig:Overall}. The setup consists of a gas / vacuum system to allow for the evacuation of the setup as well as the deployment of both the noble gas to be used in the pulsed light source as well as gas to be used in the attenuation cell. This attenuation gas can be injected in controlled amounts into the attenuation cell and then closed off from the rest of the system. All of the gas lines are connected to a common vacuum pump which is also used to evacuate the light system.

The light leaving the pulsed source is directed to a VUV monochromator (Resonance Ltd. P/N: VM200) capable of selecting individual wavelengths with nanometer precision. The light then leaves the monochromator and goes to a chamber where a light detection device under test can be mounted. For the characterization of the pulsed light source the  Hamamatsu VUV-sensitive Multi-Pixel Photon Counter (MPPC) (VUV4-MPPC P/N: S13370-6050CN) was used. VUV4-MPPC signals were amplified using two stages of 1.8 GHz wideband fixed gain amplifiers (Texas Instruments P/N: THS4304).  A Silicon photodiode (PD) (Thorlabs P/N: PDA10A2) was mounted to one of the view ports on pulsed light source, served as a reference signal.  Coincidence signals between the VUV4-MPPC and PD were digitized using a CAEN V1725 14-bit 250 MS/s waveform digitizer.


The pulsed light source assembly itself can be seen in Figure \ref{fig:Lamp} and consists of three major components: i) the spark chamber, ii) optical attenuator, and iii) the discharge electronics. 

\paragraph{\textbf{Spark Chamber}}
The spark chamber was machined from solid piece of 304 stainless steel and is meant to provide a volume into which the desired noble element gas can be injected and then via the rapid discharge of a pair of electrodes a spark can form creating light characteristic of the noble element gas. The discharge electrodes are standard 14 gauge Romex housing copper wire mounted via a casted mold and epoxied in place in the center of the spark chamber. The inner spark chamber is coated in Stycast 2850 CAT11 epoxy in order to reduce spurious breakdowns between the electrodes and the metallic housing. The epoxy was chosen because of its low outgassing and high dielectric strength. There are three optical windows for use on the spark chamber. The axial and perpendicular view ports have quartz windows (McMaster-Carr P/N: 1357T21) epoxied onto plates with TorrSeal. The plates are mounted onto the spark chamber via o-rings to allow for access into the main spark chamber or the use of alternative windows. These ports can be used for mounting auxiliary detectors or inline light sources such as lasers, LEDs, or photodiodes. The third optical window opens to the optical attenuator. This window has a 1.5 mm thick Magnesium Fluoride (MgF$_{2}$) window epoxied into a grove to allow for transmission of the VUV light (\textgreater 50\% above 120 nm) into the attenuator volume. The noble element gas source can be injected into the spark chamber via Swagelok\textregistered~VCR fittings that are epoxied in place (Torrseal) and placed as far away from the electrodes to minimize the chance of unwanted breakdown. The connection of the spark chamber to the monochromator is made via a CF 2.75 inch half nipple (Kurt Lesker P/N: HN-0275R) that was welded to the main body.

\paragraph{\textbf{Optical Attenuator}}
The optical attenuator consists of a 8 mm diameter x 10 mm long optical cavity with 1.5 mm thick MgF$_{2}$ entrance and exit windows (Edmund Optics Inc. P/N: 87-703). The purpose of the attenuator is to allow the optical absorption of the photons produced in the spark chamber via a small gas volume where the gas can be chosen to optimize the absorption of the particular light spectrum the user wishes to attenuate. The attenuator gas is brought into the optical attenuator volume via a similar Swagelok\textregistered~VCR fitting. The gas pressure and composition is controlled through a Swagelok\textregistered~VCR port located along the side of cavity. The cavity can be operated either under vacuum or with a gas composition.  Gas pressure is monitored using a 0-100 Torr Baratron\textregistered~capacitance manometer (MKS model 627B).

\paragraph{\textbf{Discharge electronics}}

The electronics utilized for the pulsed light source here are a departure from the typical flash lamp driving circuitry \cite{Emmett,Liu_2021}. As an example a standard lamp driver will utilize a transformer to generate a short high voltage pulse which initiates the discharge or an LCC resonant converter with a simmer circuit \cite{LCCResonantConverter}. Other methods \cite{birch1981coaxial} utilize a thyratron to generate a fast, high voltage pulse. The design here chooses a different approach where the two electrodes are held at some high voltage and the circuitry quickly brings one of the electrodes to ground. Approaching the circuit with this methodology allows for a standard low side MOSFET driver to be utilized and simplifies various aspects of the pulse. An overview of the driver circuitry is given in Figure \ref{fig:schematic}.

\begin{figure*}[t]
\centering
\includegraphics[width=1.0\textwidth]{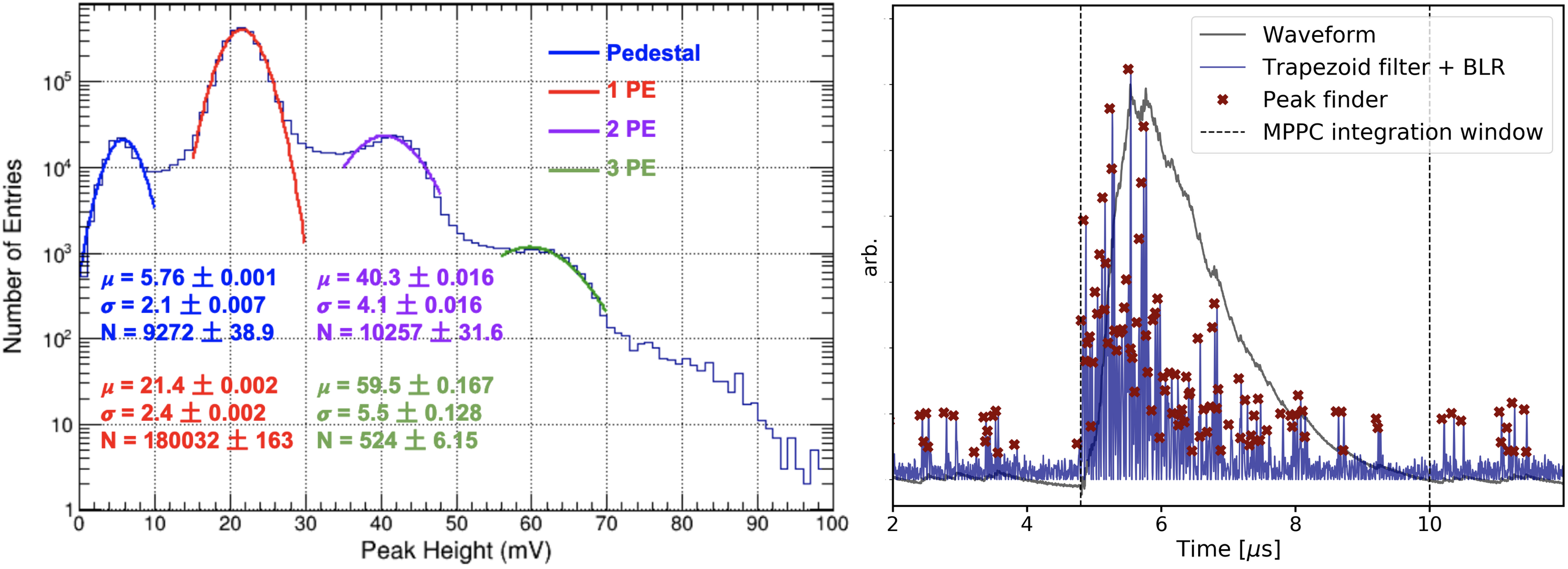}
\caption{
Left: The single photoelectron calibration of the VUV4-MPPC. The calibration was performed by considering parts of the waveform before and after the main pulse. Here the one, two, and three p.e. peaks can be seen, the s.p.e. response was then determined to be $\sim 21$~mV. Right: Summary of the analysis protocol, The raw waveform can be seen along with the filtered p.e. response which is determined by tuning a trapezoid filter and baseline restoration. A peak finder is then used with the SPE response as a threshold. The number of p.e. in a pulse is then determined by integrating the peaks in given window, where the window is defined by considering the average waveform from a run.}
\label{fig:calb_ana}
\end{figure*}

The high voltage MOSFET (Infineon Technologies P/N: IMW120R350M1HXKSA1) was selected for its fast turn on time ($7$ns at $V_{DD}=800V$) and its maximum voltage rating of $1200V$. In order to drive the MOSFET as fast as possible the gate driver (ON Semiconductor P/N: NCP81074BDR2G) is used. This MOSFET and driver combination yields a typical performance of 800V swings in $\sim$10ns. In order to control the gate driver in a consistent manner a pulse generator circuit is used. The generator is based on the LT1721, a fast quad comparator, and is broadly described in an application note \cite{williams2004signal}. However, for this application the generator was sightly modified to allow for pulses up to 10$\mu$s. The output of pulse generator is also used to generate a buffered synchronous pulse, which is often useful for data collection as it provides an external trigger which synchronizes when light from the device will be present. This mode of operating is referred to as ``Driver-mode''. The pulse generator is triggered by one of three possible inputs: an external pulse, internal clock, or a single shot which is controlled via a push button. 

A secondary mode of operation can be achieved by shorting the output of the MOSFET, which can be done by either setting the external trigger high or physically shorting the MOSFET's source and drain pins. In this mode the spark will freely happen and the time between sparks is governed by the gas pressure and the voltage applied. While this mode is not ideal, this ``RC-mode'' is commonly implemented in the lab and can provide a brighter pulse. Later sections will compare the performance of both operation modes.

\section{Methodology}\label{sec:Methodology}
In this section we describe the operation and data collecting procedures for the commissioning of the pulsed light source. To provide a basis of comparison for the performance of the pulsed light source, we describe the calibration and analysis of the VUV4-MPPC's used in the setup. 

\begin{figure*}[t]
\centering
\includegraphics[width=0.99\textwidth]{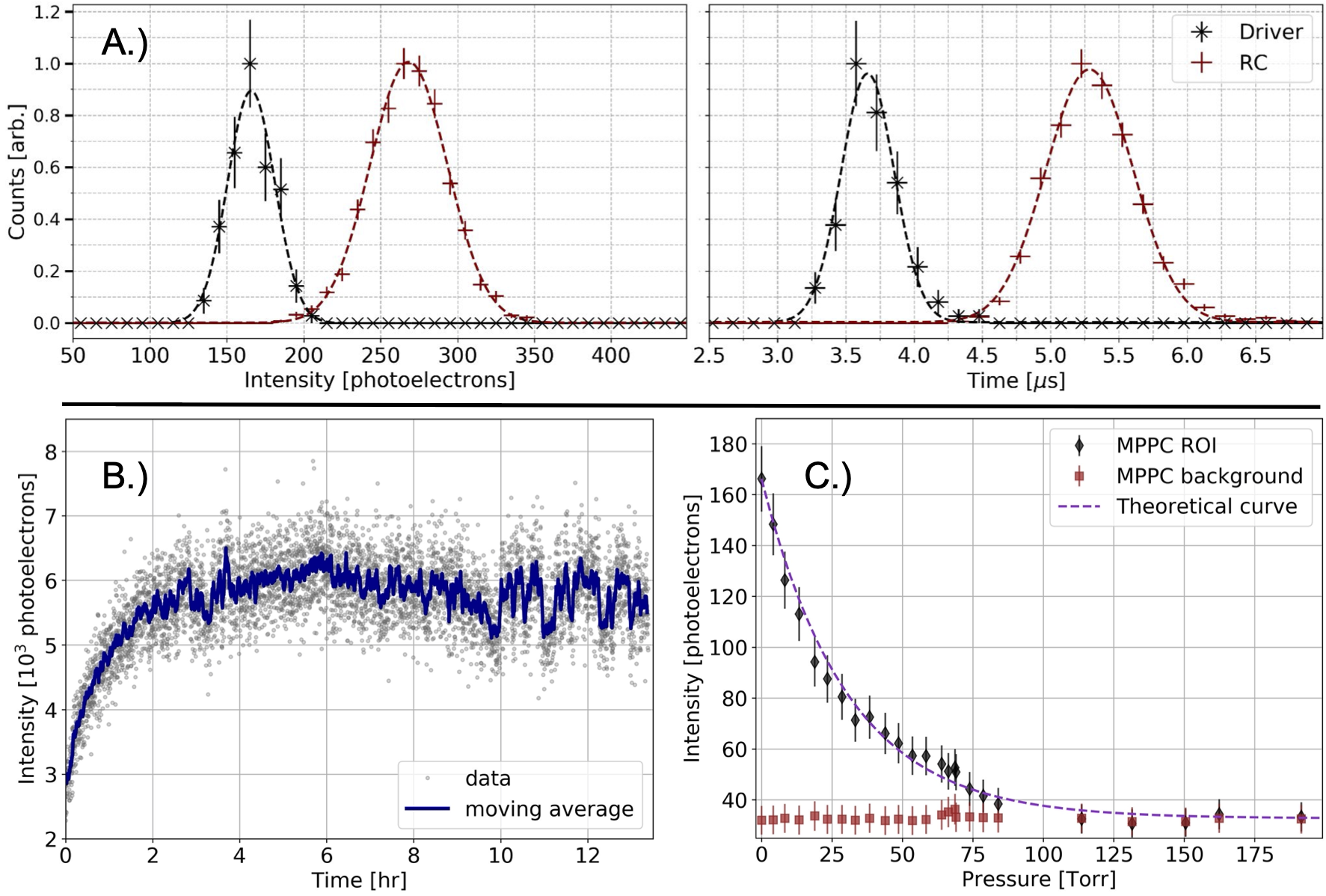}
\caption{The top panel \textbf{A} shows the comparison between the ``Driver-mode'' and ``RC-mode'', where the data used was taken from comparable runs. The analysis for the photon counting follows the methods described in section \ref{sec:calibration}. The timing profile was taken as the 10\% of the rising and falling edges as described in section \ref{sec:timing}. A Gaussian was fit to each profile and the fit parameters were used to compare the means and widths. Panel \textbf{B} shows the lamp stability in ``RC-mode'' where the lamp was running continuously over a 14 hour period. This shows the observed increase in photon yield over the first few hours of run time which is attributed to cleaning the gas as discussed in section \ref{sec:intensity}. Panel \textbf{C} is a demonstration of the capabilities of the optical attenuator to reach the single photon level. Here the momnochromator was set to 128nm and the  atteunator was filled with P-10 gas (90\% argon 10\% CH$_4$). The theoretical line is adjusted to match the starting value and include the baseline from the MPPC. The value of the CH$_4$ cross section that best fits the data here is $1\times10^{-17}$cm$^2$ per molecule. While slightly different than the global value it is within errors of other measurements.}
\label{fig:drivervsRC}
\end{figure*}

\subsection{System Operation \& Data Collection}
The spark chamber, optical attenuator, and monochromator volumes are initially evacuated to pressures $\leq 1\times 10^{-6}$~Torr. Whenever the system was not in use vacuum was maintained and the monochrometer was consistently held below this level for all measurements.

When the system was ready for operations, the spark chamber was slowly filled with a new source gas with the pressure in the chamber maintained through a series of control valves. Once the desired pressure was achieved in the spark chamber, the volume was isolated and the remainder of the gas lines were evacuated. The procedure for filling the attenuator volume is identical. All of the gases used during the subsequently described measurements have a purity rating of ultra high purity or better ($>99.999\%$) with the exception of krypton which had a purity rating of $99.998\%$. Due to this high purity and good vacuum achieved in the system, the gas was used directly from the bottle with out further purification.


For all measurements and calibrations shown the operating voltage of the VUV4-MPPC was set to -55~V. The waveforms from the VUV4-MPPC and PD were recorded on the digitizer and the digitizer was triggered by the signal from the PD. This readout and triggering scheme was used for all measurements to allow for consistency in the comparisons between the ``Driver-mode'' and ``RC-mode'' operations of the pulsed light source. Wavelength scans were performed by running the lamp in driver mode. The wavelength of the monochormeter was controlled via an Arduino microcontroller \cite{arduino} which would toggle TTL signal indicating it was moving. For wavelength scans this signal was also digitized and used during the analysis to correlate sets of pulses with a given wavelength. In all cases the wavelength scans were performed with a step size of 0.5~nm over the range 100-200~nm.


The monochromator has two sets of optical slits, located at the entrance and exit. These optical slits allow you to further adjust the intensity of the light leaving the monochrometer as well as the wavelength resolution. During the wavelength scans, these optical slits had to be adjusted due to the relative intensity of the specific wavelengths produced by the source gas as well as the varying quantum efficiency of the VUV4-MPPC. Whenever an adjustment was made, the scan was retaken over the range of relevant wavelengths in order to allow for the data to cross-compared and ultimately compiled into a single relative intensity shown for each source gas.

Optical attenuation measurements started by setting the monochrmator to 128~nm. The source cavity was then filled with argon gas then the gas lines were evacuated. Once evacuated, a set of measurements were taken to indicate the initial, unattenuated intensity.. A small amount of P-10 gas was then added to the optical attenuation cavity and another set of measurements were performed. This process was repeated in $\sim4$~Torr steps until the VUV4-MPPC signal disappeared. Measurements were then taken at higher pressures after the signal had ceased being visible in order to confirm the behavior of the optical attenuator.

\subsection{\label{sec:calibration}Calibration}

In order to accurately measure the number of photons produced by the pulsed light source, data from the VUV4-MPPC was used to determine the single photoelectron (SPE) response. For every waveform recorded during the calibration run a ``pre-window'' and ''post-window'' time frame was used which surrounds the typical pulse seen from the PD attached to the spark chamber. The ``pre-window used'' was defined relative to the start of the digitizer readout ($t=0 ~\mu$s) as $10 ~\mu$s $\leq t \leq 1050 ~\mu$s and the ``post-window'' is defined as $3000 ~\mu$s $\leq t \leq 4500 ~\mu$s. The monochromator was set to 100~nm and the digitizer was allowed to auto trigger until a sufficient number of pulses were collected.

The VUV4-MPPC analysis methods implemented here broadly follow the steps described in G. Knoll's text for VUV4-MPPC's\cite{knoll2010radiation}. The raw VUV4-MPPC waveform was inverted (negative to positive) and the baseline offset removed. This was done to aid in the progressing analysis steps. A trapezoidal filter is then applied to the waveform, which produces the finite impulse response of the VUV4-MPPC \cite{jordanov1994digital}. However, in regions with a high number of pileup events, the recursive trapezoidal filter overshoots the baseline and a further correction is needed. These overshoots are restored using an implementation of the standard SNIP algorithm \cite{ryan1988snip, morhavc2008peak}. With the baseline corrected and the electronics response removed, photoelectrons are then identified with a peak finding algorithm. All of the steps above are stored and referenced for the different analysis.

The SPE response of the VUV4-MPPC can be determined through analysis of dark pulses using the pulse height of the trapezodial filtered events. Histograming all the pulse heights yields the histogram shown on the right hand side of Figure \ref{fig:calb_ana}. The fit to this histogram provides both the SPE scale of 21.4~mV and a linear extrapolation using the three SPE peaks fitted provide the basis of the estimate of the number of PE used in later analysis.

Calibration of the VUV monochromator was performed using known strong neutral atomic spectral lines in carbon, nitrogen, and xenon between 119 - 193~nm. Spectra of research grade nitrogen and xenon were acquired and calibrated against NIST Standard Reference Database 108 \ref{NIST}.  Neutral carbon atomic spectral lines, observed in all spectra acquired, are suspected to be due to interaction of the gas discharge with the epoxy coated chamber.\\

\section{\label{sec:results}Results and Discussion}
Four results can be extracted from the data taken with the pulse light source that are of importance to its use as a calibration source. These include the timing profile of the light pulse created by the spark, the intensity of the light produced, the stability of the light source over time, the ability of the optical attenuator to achieve low photon flux, and finally the observed emission spectra for various source gases. In this section, we will present and discuss each of these aspects measured for this device.

\subsection{\label{sec:timing}Timing Profile}
The timing profile of the light created during the pulse can be extracted from the digitized waveforms of the VUV4-MPPC's. For this measurement both the ``Driver-mode'' and ``RC-mode'' of discharge were used with an argon source gas and the monochrometer set for 128~nm. We extract the timing profile on a pulse-by-pulse basis using the inverted raw VUV4-MPPC waveform (and example is shown in Figure \ref{fig:calb_ana} in gray). The peak of the waveform is found and then the two times corresponding to 10\% of the peak are found. The time difference of these points is then stored as the pulse width. The distribution of pulse widths is then plotted in Figure \ref{fig:drivervsRC} \textbf{A} and the distribution is fit with a Gaussian function.

In RC-mode, the pulse width was found to have a 5.27$\pm$0.748~$\mu$s (FWHM) while in Driver-mode the pulse width is significantly lower at 3.65$\pm$0.425~$\mu$s. This timing width of the light generated by the spark is longer than the typical ``fast'' time constants of noble element scintillation coming from the singlet state and the ``late'' component coming from the triplet state. While the timing profile is longer than anticipated its stability will still allow useful calibration of sensors.


\subsection{\label{sec:intensity}Relative Light Intensity}

The shot-by-shot photon intensity is calculated using an integration window of an averaged pulse for a given set of measurements. The size of the integration window is determined by considering the start and end of a set of averaged waveforms. Once the integration window is found, the peaks of the trapezoidal filtered + baseline corrected pulses are summed and stored as the pulse intensity. The SPE calibration is then used to translate this to an intensity in photoelectrons. The distribution of intensities are then plotted in a histogram and fit with a Gaussian function, as shown in the left hand side of Figure \ref{fig:drivervsRC} \textbf{A}

In RC-mode, the VUV photon intensity was found have a mean of 268 p.e with an intensity spread of 59 p.e. (FWHM). In driver-mode the VUV photon intensity was found to be lower at 166 p.e with a shot-by-shot intensity spread of 36 p.e. (FWHM).  Correcting for the quantum efficiency and fill fraction of the VUV4-MPPC (14.24\%), this gives a yield of 1883 p.e and 1164 p.e for operation in RC and Driver modes, respectively. The overall pulse to pulse systematic uncertainty is 10\% over all wavelengths and gases. This uncertainty is determined by identifying various comparable peaks in each spectrum and assessing the variance of the pulse width.

\subsection{\label{sec:stabilityh}Stability}
An important performance metric is the time-dependence stability of the pulsed light source.  To evaluate this, the pulse light source was filled with 14 Torr of Argon and operated at a trigger rate of 10~Hz for over 13 hours in RC mode. The monochromator was set to 128~nm and the slits were further opened to allow the full 128~nm peak to be used. Figure \ref{fig:drivervsRC} \textbf{B} shows the time-dependence of 128 nm photons over the 13 hour measurement.  The intensity was found to rapidly increase over a $\sim$ 4 hour time frame, finally stability at double the initial intensity. This effect was observed numerous times and is suspected to be due to the discharge cleaning residual impurities in the gas. While not a standard method to clean gas it has been demonstrated to have this effect\cite{pokachalov1993spark}.

\begin{figure*}[t]
\centering
\includegraphics[width=0.99\textwidth]{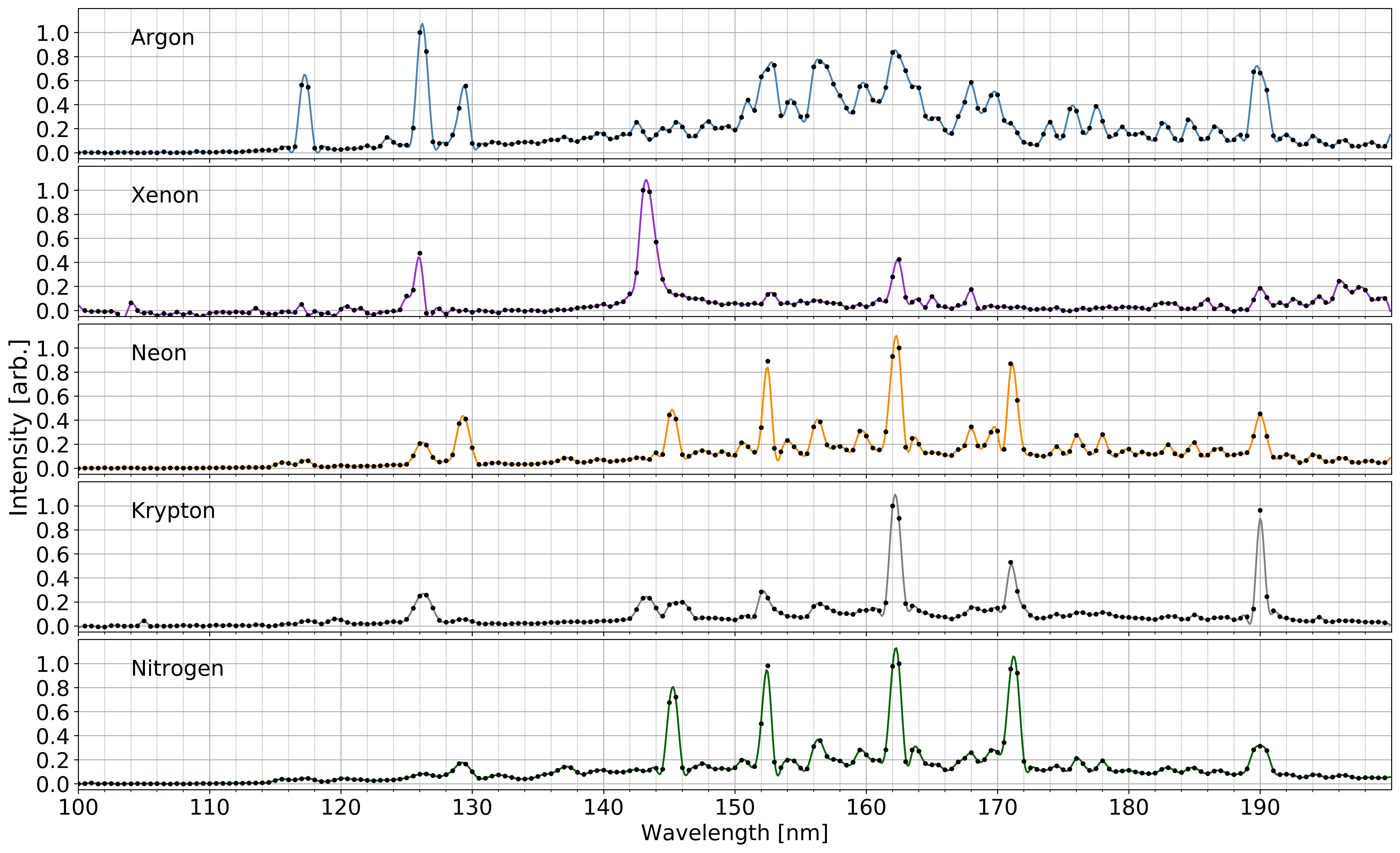}
\caption{Emission spectra taken with the source using ``Driver-mode'' in nitrogen, argon, neon, krypton, and xenon gas. The line connecting the points is only drawn to guide the eye. Due to the pressures of the gas the emission spectra is dominated by atomic transitions. A notable feature is the common lines attributed to carbon which are apparent in each spectra, particularly lines at 119, 127, 133, 156, and 174 nm. }
\label{fig:spectra}
\end{figure*}

\subsection{\label{sec:Attenuation}Optical Attenuation}

A unique approach to this VUV pulsed light source is the ability to control the intensity of the VUV light \emph{independently} from the pulsed light source itself.  This allows for intensity control without disrupting the discharge, which would alter the timing and spectral properties.  To do this, P-10 gas is introduced into the optical attenuation cavity mounted in front of the pulsed light source.  Aforementioned, CH$_4$ in P-10 strongly absorbs VUV light below $\sim$ 145 nm thus makes is a convenient attenuation gas.  Figure \ref{fig:drivervsRC} \textbf{C} shows the intensity of 128 nm photons as a function of P-10 gas pressure in the optical attenuator.  By digitizing 1 $\mu$s before the trigger, the VUV4-MPPC background could be determined by measuring the photoelectron intensity before the VUV pulse on a event-by-event basis.  The optical attenuator was able to reach SPE levels between 70-90 Torr, which matches the 1.5\% to 0.5\% transmission predicted by theory.

\subsection{\label{sec:spectra}Emission Spectra}

Emission spectra of nitrogen, argon, neon, krypton, xenon were acquired from 100 - 200 nm using 0.5 nm steps of the monochrometer. In addition to pure gases, the emission spectra of air was also acquired over the same wavelength range.  Figure \ref{fig:spectra} shows the emission spectra of each gas measured.  Below $\sim$115 nm, the spectra are cut off due to the loss of optical transmission of the MgF$_{2}$ windows used on the attenuator and pulsed light source. The emission spectra are characteristic of strong atomic emission lines from neutral elements.  Weak neutral carbon atomic emission lines at 119, 127, 133, 156, and 174 nm are visible in all spectra and are assumed to be due to either carbon in the electrodes or interaction of discharge with the epoxy cell liner.

Of interest to the HEP community is the emission spectra of the noble elements, particularly argon. In the emission range relevant for liquid argon (LAr) scintillation, argon emission spectrum is characterized by three neutral (Ar I) emission lines at 117.7, 126.7, and 129.8 nm \cite{}.  The argon dimer Ar$_{2}^{*}$ emission at 128 nm, which is characteristic for LAr scintillation, is noticeably absent in the spectra since the operating conditions are unfavorable for Ar$_{2}$ formation.

\section{\label{sec:conclusion}Conclusion }

A novel pulsed VUV light source with adjustable intensity is presented here.  The light source exhibits acceptable timing performance with pulse width between 3.65-5.27 $\mu$s, while longer than LAr scintillation it is still expected to be useful for LAr detector research for HEP.  At a 128 nm light yield of $\mathcal{O}(1500)$ p.e. which is attenuatable down to single p.e. levels for devices under test. The device has been shown to be stable, following a gas clean-up phase, over a period of many hours. This device is available for use at Oak Ridge National Labs photon detector development platform. \\

\section{\label{sec:acknowledgement}acknowledgement }

The authors would like to thank Scott Eichmann from the University of Texas Arlington machine shop for fabricating the lamp. Calibration and testing of the device was performed at Oak Ridge National Labs which is a DOE Office of Science User Facility. This material is based upon work supported by the U.S. Department of Energy, Office of Science, Office of High Energy Physics Award Number DE-SC0020065.

\bibliography{ArLamp}

\begin{thebibliography}{16}%
\makeatletter
\providecommand \@ifxundefined [1]{%
 \@ifx{#1\undefined}
}%
\providecommand \@ifnum [1]{%
 \ifnum #1\expandafter \@firstoftwo
 \else \expandafter \@secondoftwo
 \fi
}%
\providecommand \@ifx [1]{%
 \ifx #1\expandafter \@firstoftwo
 \else \expandafter \@secondoftwo
 \fi
}%
\providecommand \natexlab [1]{#1}%
\providecommand \enquote  [1]{``#1''}%
\providecommand \bibnamefont  [1]{#1}%
\providecommand \bibfnamefont [1]{#1}%
\providecommand \citenamefont [1]{#1}%
\providecommand \href@noop [0]{\@secondoftwo}%
\providecommand \href [0]{\begingroup \@sanitize@url \@href}%
\providecommand \@href[1]{\@@startlink{#1}\@@href}%
\providecommand \@@href[1]{\endgroup#1\@@endlink}%
\providecommand \@sanitize@url [0]{\catcode `\\12\catcode `\$12\catcode
  `\&12\catcode `\#12\catcode `\^12\catcode `\_12\catcode `\%12\relax}%
\providecommand \@@startlink[1]{}%
\providecommand \@@endlink[0]{}%
\providecommand \url  [0]{\begingroup\@sanitize@url \@url }%
\providecommand \@url [1]{\endgroup\@href {#1}{\urlprefix }}%
\providecommand \urlprefix  [0]{URL }%
\providecommand \Eprint [0]{\href }%
\providecommand \doibase [0]{https://doi.org/}%
\providecommand \selectlanguage [0]{\@gobble}%
\providecommand \bibinfo  [0]{\@secondoftwo}%
\providecommand \bibfield  [0]{\@secondoftwo}%
\providecommand \translation [1]{[#1]}%
\providecommand \BibitemOpen [0]{}%
\providecommand \bibitemStop [0]{}%
\providecommand \bibitemNoStop [0]{.\EOS\space}%
\providecommand \EOS [0]{\spacefactor3000\relax}%
\providecommand \BibitemShut  [1]{\csname bibitem#1\endcsname}%
\let\auto@bib@innerbib\@empty
\bibitem [{\citenamefont {Zheng}\ \emph {et~al.}(2020)\citenamefont {Zheng},
  \citenamefont {Jia},\ and\ \citenamefont {Huang}}]{Zheng2020}%
  \BibitemOpen
  \bibfield  {author} {\bibinfo {author} {\bibfnamefont {W.}~\bibnamefont
  {Zheng}}, \bibinfo {author} {\bibfnamefont {L.}~\bibnamefont {Jia}},\ and\
  \bibinfo {author} {\bibfnamefont {F.}~\bibnamefont {Huang}},\ }\href
  {https://doi.org/10.1016/j.isci.2020.101145} {\bibfield  {journal} {\bibinfo
  {journal} {iScience}\ }\textbf {\bibinfo {volume} {23}},\ \bibinfo {pages}
  {101145} (\bibinfo {year} {2020})},\ \bibinfo {note}
  {32446223[pmid]}\BibitemShut {NoStop}%
\bibitem [{\citenamefont {Majumdar}\ and\ \citenamefont
  {Mavrokoridis}(2021)}]{app11062455}%
  \BibitemOpen
  \bibfield  {author} {\bibinfo {author} {\bibfnamefont {K.}~\bibnamefont
  {Majumdar}}\ and\ \bibinfo {author} {\bibfnamefont {K.}~\bibnamefont
  {Mavrokoridis}},\ }\bibfield  {journal} {\bibinfo  {journal} {Applied
  Sciences}\ }\textbf {\bibinfo {volume} {11}},\ \href
  {https://doi.org/10.3390/app11062455} {10.3390/app11062455} (\bibinfo {year}
  {2021})\BibitemShut {NoStop}%
\bibitem [{\citenamefont {Baudis}(2014)}]{BAUDIS201450}%
  \BibitemOpen
  \bibfield  {author} {\bibinfo {author} {\bibfnamefont {L.}~\bibnamefont
  {Baudis}},\ }\href
  {https://doi.org/https://doi.org/10.1016/j.dark.2014.07.001} {\bibfield
  {journal} {\bibinfo  {journal} {Physics of the Dark Universe}\ }\textbf
  {\bibinfo {volume} {4}},\ \bibinfo {pages} {50} (\bibinfo {year} {2014})},\
  \bibinfo {note} {dARK TAUP2013}\BibitemShut {NoStop}%
\bibitem [{deu()}]{deutlamp}%
  \BibitemOpen
  \href@noop {} {\bibinfo {title} {Vacuum ultraviolet deuterium light
  source}},\ \bibinfo {howpublished}
  {\url{https://mcphersoninc.com/pdf/634.pdf}},\ \bibinfo {note} {accessed:
  09-30-2021}\BibitemShut {NoStop}%
\bibitem [{xen()}]{xenonflashlam}%
  \BibitemOpen
  \href@noop {} {\bibinfo {title} {Xenon flash lamps}},\ \bibinfo
  {howpublished}
  {\url{https://www.hamamatsu.com/resources/pdf/etd/Xe-F_TLS1022E.pdf}},\
  \bibinfo {note} {accessed: 09-30-2021}\BibitemShut {NoStop}%
\bibitem [{\citenamefont {Williams}(2004)}]{williams2004signal}%
  \BibitemOpen
  \bibfield  {author} {\bibinfo {author} {\bibfnamefont {J.}~\bibnamefont
  {Williams}},\ }\href@noop {} {\bibfield  {journal} {\bibinfo  {journal}
  {Linear Technology Corporation, Application Note}\ }\textbf {\bibinfo
  {volume} {98}},\ \bibinfo {pages} {20} (\bibinfo {year} {2004})}\BibitemShut
  {NoStop}%
\bibitem [{\citenamefont {Markiewicz}(1966)}]{Emmett}%
  \BibitemOpen
  \bibfield  {author} {\bibinfo {author} {\bibfnamefont {J.}~\bibnamefont
  {Markiewicz}, \bibfnamefont {J.~;~Emmett}},\ }\href@noop {} {\bibfield
  {journal} {\bibinfo  {journal} {IEEE Journal of Quantum Electronics}\
  }\textbf {\bibinfo {volume} {QE-2}},\ \bibinfo {pages} {707} (\bibinfo {year}
  {1966})}\BibitemShut {NoStop}%
\bibitem [{\citenamefont {Liu}\ \emph {et~al.}(2021)\citenamefont {Liu},
  \citenamefont {Ma},\ and\ \citenamefont {Zhang}}]{Liu_2021}%
  \BibitemOpen
  \bibfield  {author} {\bibinfo {author} {\bibfnamefont {D.}~\bibnamefont
  {Liu}}, \bibinfo {author} {\bibfnamefont {M.}~\bibnamefont {Ma}},\ and\
  \bibinfo {author} {\bibfnamefont {Y.}~\bibnamefont {Zhang}},\ }\href
  {https://doi.org/10.1088/1742-6596/1754/1/012016} {\bibfield  {journal}
  {\bibinfo  {journal} {Journal of Physics: Conference Series}\ }\textbf
  {\bibinfo {volume} {1754}},\ \bibinfo {pages} {012016} (\bibinfo {year}
  {2021})}\BibitemShut {NoStop}%
\bibitem [{\citenamefont {Song}\ \emph {et~al.}(2019)\citenamefont {Song},
  \citenamefont {Cho}, \citenamefont {Park}, \citenamefont {Park},
  \citenamefont {Jeong},\ and\ \citenamefont {Ryoo}}]{LCCResonantConverter}%
  \BibitemOpen
  \bibfield  {author} {\bibinfo {author} {\bibfnamefont {S.-H.}\ \bibnamefont
  {Song}}, \bibinfo {author} {\bibfnamefont {C.-G.}\ \bibnamefont {Cho}},
  \bibinfo {author} {\bibfnamefont {S.-M.}\ \bibnamefont {Park}}, \bibinfo
  {author} {\bibfnamefont {H.-I.}\ \bibnamefont {Park}}, \bibinfo {author}
  {\bibfnamefont {W.-C.}\ \bibnamefont {Jeong}},\ and\ \bibinfo {author}
  {\bibfnamefont {H.}~\bibnamefont {Ryoo}},\ }\href
  {https://doi.org/10.1109/TDEI.2019.007696} {\bibfield  {journal} {\bibinfo
  {journal} {IEEE Transactions on Dielectrics and Electrical Insulation}\
  }\textbf {\bibinfo {volume} {26}},\ \bibinfo {pages} {484} (\bibinfo {year}
  {2019})}\BibitemShut {NoStop}%
\bibitem [{\citenamefont {Birch}\ and\ \citenamefont
  {Imhof}(1981)}]{birch1981coaxial}%
  \BibitemOpen
  \bibfield  {author} {\bibinfo {author} {\bibfnamefont {D.}~\bibnamefont
  {Birch}}\ and\ \bibinfo {author} {\bibfnamefont {R.}~\bibnamefont {Imhof}},\
  }\href@noop {} {\bibfield  {journal} {\bibinfo  {journal} {Review of
  Scientific Instruments}\ }\textbf {\bibinfo {volume} {52}},\ \bibinfo {pages}
  {1206} (\bibinfo {year} {1981})}\BibitemShut {NoStop}%
\bibitem [{ard()}]{arduino}%
  \BibitemOpen
  \href@noop {} {\bibinfo {title} {Arduino}},\ \bibinfo {howpublished}
  {\url{https://www.arduino.cc/}},\ \bibinfo {note} {accessed:
  2021-09-30}\BibitemShut {NoStop}%
\bibitem [{\citenamefont {Knoll}(2010)}]{knoll2010radiation}%
  \BibitemOpen
  \bibfield  {author} {\bibinfo {author} {\bibfnamefont {G.~F.}\ \bibnamefont
  {Knoll}},\ }\href@noop {} {\emph {\bibinfo {title} {Radiation detection and
  measurement}}}\ (\bibinfo  {publisher} {John Wiley \& Sons},\ \bibinfo {year}
  {2010})\BibitemShut {NoStop}%
\bibitem [{\citenamefont {Jordanov}\ and\ \citenamefont
  {Knoll}(1994)}]{jordanov1994digital}%
  \BibitemOpen
  \bibfield  {author} {\bibinfo {author} {\bibfnamefont {V.~T.}\ \bibnamefont
  {Jordanov}}\ and\ \bibinfo {author} {\bibfnamefont {G.~F.}\ \bibnamefont
  {Knoll}},\ }\href@noop {} {\bibfield  {journal} {\bibinfo  {journal} {Nuclear
  Instruments and Methods in Physics Research Section A: Accelerators,
  Spectrometers, Detectors and Associated Equipment}\ }\textbf {\bibinfo
  {volume} {345}},\ \bibinfo {pages} {337} (\bibinfo {year}
  {1994})}\BibitemShut {NoStop}%
\bibitem [{\citenamefont {Ryan}\ \emph {et~al.}(1988)\citenamefont {Ryan},
  \citenamefont {Clayton}, \citenamefont {Griffin}, \citenamefont {Sie},\ and\
  \citenamefont {Cousens}}]{ryan1988snip}%
  \BibitemOpen
  \bibfield  {author} {\bibinfo {author} {\bibfnamefont {C.}~\bibnamefont
  {Ryan}}, \bibinfo {author} {\bibfnamefont {E.}~\bibnamefont {Clayton}},
  \bibinfo {author} {\bibfnamefont {W.}~\bibnamefont {Griffin}}, \bibinfo
  {author} {\bibfnamefont {S.}~\bibnamefont {Sie}},\ and\ \bibinfo {author}
  {\bibfnamefont {D.}~\bibnamefont {Cousens}},\ }\href@noop {} {\bibfield
  {journal} {\bibinfo  {journal} {Nuclear Instruments and Methods in Physics
  Research Section B: Beam Interactions with Materials and Atoms}\ }\textbf
  {\bibinfo {volume} {34}},\ \bibinfo {pages} {396} (\bibinfo {year}
  {1988})}\BibitemShut {NoStop}%
\bibitem [{\citenamefont {Morh{\'a}{\v{c}}}\ and\ \citenamefont
  {Matou{\v{s}}ek}(2008)}]{morhavc2008peak}%
  \BibitemOpen
  \bibfield  {author} {\bibinfo {author} {\bibfnamefont {M.}~\bibnamefont
  {Morh{\'a}{\v{c}}}}\ and\ \bibinfo {author} {\bibfnamefont {V.}~\bibnamefont
  {Matou{\v{s}}ek}},\ }\href@noop {} {\bibfield  {journal} {\bibinfo  {journal}
  {Applied spectroscopy}\ }\textbf {\bibinfo {volume} {62}},\ \bibinfo {pages}
  {91} (\bibinfo {year} {2008})}\BibitemShut {NoStop}%
\bibitem [{\citenamefont {Pokachalov}\ \emph {et~al.}(1993)\citenamefont
  {Pokachalov}, \citenamefont {Kirsanov}, \citenamefont {Kruglov},\ and\
  \citenamefont {Obodovski}}]{pokachalov1993spark}%
  \BibitemOpen
  \bibfield  {author} {\bibinfo {author} {\bibfnamefont {S.}~\bibnamefont
  {Pokachalov}}, \bibinfo {author} {\bibfnamefont {M.}~\bibnamefont
  {Kirsanov}}, \bibinfo {author} {\bibfnamefont {A.}~\bibnamefont {Kruglov}},\
  and\ \bibinfo {author} {\bibfnamefont {I.}~\bibnamefont {Obodovski}},\
  }\href@noop {} {\bibfield  {journal} {\bibinfo  {journal} {Nuclear
  Instruments and Methods in Physics Research Section A: Accelerators,
  Spectrometers, Detectors and Associated Equipment}\ }\textbf {\bibinfo
  {volume} {327}},\ \bibinfo {pages} {159} (\bibinfo {year}
  {1993})}\BibitemShut {NoStop}%
\end{thebibliography}%

\end{document}